\documentclass[12pt]{iopart}
\expandafter\let\csname equation*\endcsname\relax
\expandafter\let\csname endequation*\endcsname\relax
\usepackage{amsmath,amssymb,amsfonts}
\usepackage{cite}
\usepackage{algorithmic}
\usepackage{textcomp}
\usepackage{xcolor}
\usepackage{hyperref}
\usepackage{graphicx} % Required for inserting images
\usepackage{makecell}
\usepackage{multirow}
\usepackage{bm}
 \usepackage{verbatim}

\begin{document}

\title[PDS-MAR for intraoperative CBCT images with guidewires]{PDS-MAR: a fine-grained Projection-Domain Segmentation-based Metal Artifact Reduction method for intraoperative CBCT images with guidewires}
\author{Tianling Lyu$^1$, Zhan Wu$^2$, Gege Ma$^1$, Chen Jiang$^1$, Xinyun Zhong$^2$, Yan Xi$^3$, Yang Chen$^{2,4,*}$ and Wentao Zhu$^{1,*}$}
\address{$^1$ Research Center for Augmented Intelligence, Zhejiang Lab, Hangzhou, China}
\address{$^2$ Laboratory of Imaging Science and Technology, Southeast University, Nanjing, China}
\address{$^3$ First-Imaging Tech., Shanghai, China}
\address{$^4$ Jiangsu Provincial Joint International Research Laboratory of Medical Information Processing, Southeast University, Nanjing, China.}
\address{$^*$ Authors to whom any correspondence should be addressed. }
\ead{\url{wentao.zhu@zhejianglab.com} and \url{chenyang.list@seu.edu.cn}}
\vspace{10pt}

\begin{abstract}
Since the invention of modern CT systems, metal artifacts have been a persistent problem. Due to increased scattering, amplified noise, and insufficient data collection, it is more difficult to suppress metal artifacts in cone-beam CT, limiting its use in human- and robot-assisted spine surgeries where metallic guidewires and screws are commonly used. In this paper, we demonstrate that conventional image-domain segmentation-based MAR methods are unable to eliminate metal artifacts for intraoperative CBCT images with guidewires. To solve this problem, we present a fine-grained projection-domain segmentation-based MAR method termed PDS-MAR, in which metal traces are augmented and segmented in the projection domain before being inpainted using triangular interpolation. In addition, a metal reconstruction phase is proposed to restore metal areas in the image domain. The digital phantom study and real CBCT data study demonstrate that the proposed algorithm achieves significantly better artifact suppression than other comparing methods and has the potential to advance the use of intraoperative CBCT imaging in clinical spine surgeries. 

\end{abstract}

\section{Introduction}
\label{sec:intro}
Minimally Invasive Spine Surgery (MISS) is a surgical procedure designed to stabilize the vertebral bones and spinal joints or relieve pressure on the spinal nerves caused by conditions such as herniated discs, lumbar spinal stenosis, spinal tumors, etc. Compared to conventional open spine surgery, MISS results in a significantly shorter surgical time, a quicker patient recovery, and a reduced risk of infection and postoperative discomfort \cite{yoon2019evolution, goldstein2016perioperative, wang2017systematic}. In recent decades, it has gained popularity as a treatment for spinal disorders \cite{spetzger2013past, vaishnav2019current, park2020minimally}. Depending on the condition of the patient, instrumentation like rods and screws may be necessary to stabilize the spine. To minimize skin incisions, guidewires are inserted through the skin and into the spinal vertebrae along the instrumentation's desired trajectories. Typically, this phase is navigated using fluoroscope images from 2D C-arm systems \cite{kim2008use, tjardes2010image, kraus2013image}. However, these systems must be manually adjusted to a specific angle to get a better view, which makes them incompatible with modern robotic-assisted surgery systems. Recent studies have shown that intraoperative 3-D imaging systems are more dependable than fluoroscope systems for guidewire and screw navigation, and may reduce surgical complications \cite{vaishnav2019current, overley2017navigation}. 3-D imaging systems can also provide better image navigation for robotic-assisted surgery systems \cite{park2020minimally, kochanski2019image}. Intraoperative 3-D imaging systems are promised to become more and more important in MISS with the thriving of robotic-assisted surgery systems. 

Cone-beam computed tomography (CBCT) is one of the most widely used intraoperative 3-D imaging modalities that has already been adopted in orthopedic surgery \cite{tonetti2020role, tkatschenko2020navigated, siewerdsen2020cone}. Compared to the heavy-weighted multi-slice CT (MSCT) commonly applied for diagnosis, CBCTs are usually light-weighted mobile systems producing much-lowered irradiation doses, making them the optimal imaging modality for intraoperative applications. However, metal artifacts introduced by metallic guidewires can significantly degrade images and impair the physician's ability to determine wire location \cite{schafer2020technology}. Metal artifact has long been the cause of image quality degradation in all CT systems, and researchers have been addressing this issue for more than four decades \cite{gjesteby2016metal}. These artifacts are more difficult to suppress in CBCT due to increased scattering, magnified noise, and insufficient data collection. 

\begin{figure}
    \centering
    \includegraphics[width=0.7\textwidth]{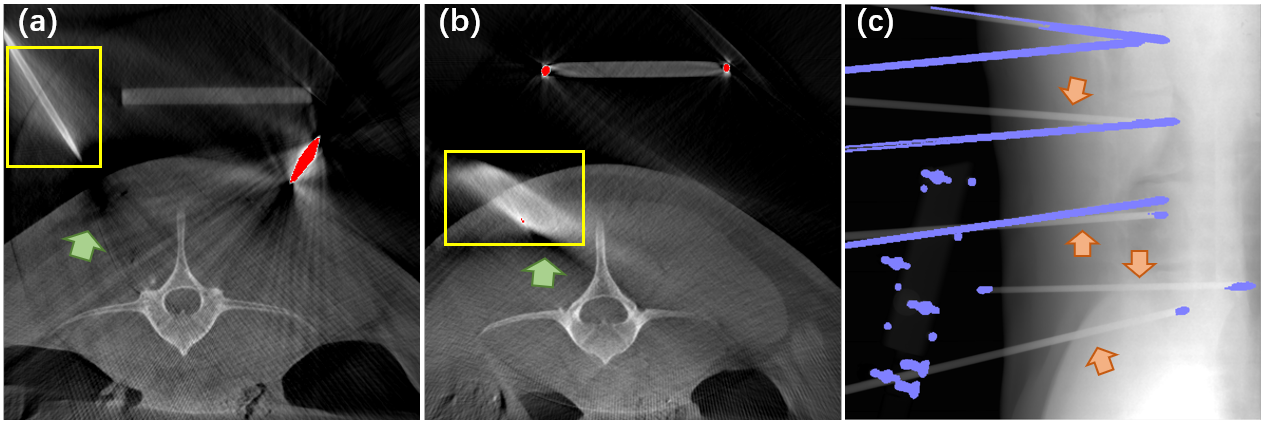}
    \caption{An example from a commercial CBCT system. (a) and (b) are two slices reconstructed using FDK algorithm, red regions show metal objects segmented with a threshold of 2000HU. Images are displayed under a window of (C=500HU, W=3000HU); (c) is the corresponding projection data at the first view, violet regions show the segmented metal traces. }
    \label{fig:met_examp}
\end{figure}

The majority of current metal artifact reduction (MAR) methods involve two steps: segmentation and interpolation. The segmentation step distinguishes the metal-affected region from the unaffected region in the projection domain while the interpolation step inpaints the metal-affected region (metal trace) to generate projection data without metal. Recent related literature focuses primarily on the interpolation step. The metal-trace has been inpainted with linear interpolation (LI) \cite{kalender1987reduction}, prior-image-based interpolation \cite{Meyer2010, zhang2013hybrid, zhang2014metal, Wang2013}, and convolutional neural networks \cite{zhang2018convolutional, Lin2019, Lyu2020, yu2020deep, wang2021dan, Yu2021}. These methods assume that metal can be readily segmented in the image domain, and that metal traces can be extracted via forward-projecting metal masks. However, this strategy is not always effective with CBCT. Referring to Fig. \ref{fig:met_examp} as an example, Some metallic objects are successfully segmented (red regions in Fig. \ref{fig:met_examp}(a) and Fig. \ref{fig:met_examp}(b)) whereas the others are hard to segment completely with a threshold (yellow rectangles) due to beam-hardening, cone-beam artifact, limited angle artifact and other image degradations. These metallic guidewires also introduce metal artifacts to the images (green arrows). Fig. \ref{fig:met_examp}(c) depicts a projection view corresponding to the failed segmentation in the image domain, which shows the absence of metal traces (orange arrows). To overcome the segmentation inaccuracy problem, researchers have also studied projection domain metal segmentation methods \cite{meilinger2011projective, hegazy2019u, gottschalk2021view, zhu2021ct}. However, data in the projection domain does not exhibit uniformity in fragments, making segmentation even more difficult. Even if one can obtain accurate metal traces, there is still a problem to retain image-domain metal locations from projection-domain masks.

To improve the quality of intraoperative imaging for MISS and robotic-assisted MISS, a fine-grained Projection-Domain Segmentation-based MAR algorithm (PDS-MAR) targeting metallic guidewires in intraoperative CBCT images is proposed. In the proposed algorithm, there are three distinct stages. First, guidewires are segmented using tubular enhancement filtering in the projection domain. The metal-free images are then reconstructed utilizing triangulation-based interpolation and the FDK algorithm. Finally, backprojection of multiplicative forms is used to reconstruct image-domain metal masks. 

The contributions of this work are summarized as follows. 
\begin{itemize}
    \item We demonstrate that it is impossible for image-domain segmentation-based MAR methods to eliminate CBCT metal artifacts in certain MISS situations, highlighting the significance of projection-domain metal segmentation. 
    \item We present a novel metal artifact reduction algorithm termed PDS-MAR for intraoperative CBCT with guidewires, which incorporates tubular enhancement filtering for metal segmentation and Delauney triangulation for metal trace inpainting. 
    \item We pose the problem of reconstructing image-domain metal masks from projection-domain metal traces and propose a method based on multiplicative-form backprojection as a phase in PDS-MAR to partially solve the problem. 
\end{itemize}

The rest of this article is described as follows. Section 2 illustrates the details of our PDS-MAR algorithm. Section 3 shows the experimental setting and results. Sections 4 and 5 present the discussion and conclusion of this study, respectively.

\section{Method}
\label{sec:method}
\begin{figure}
    \centering
    \includegraphics[width=\textwidth]{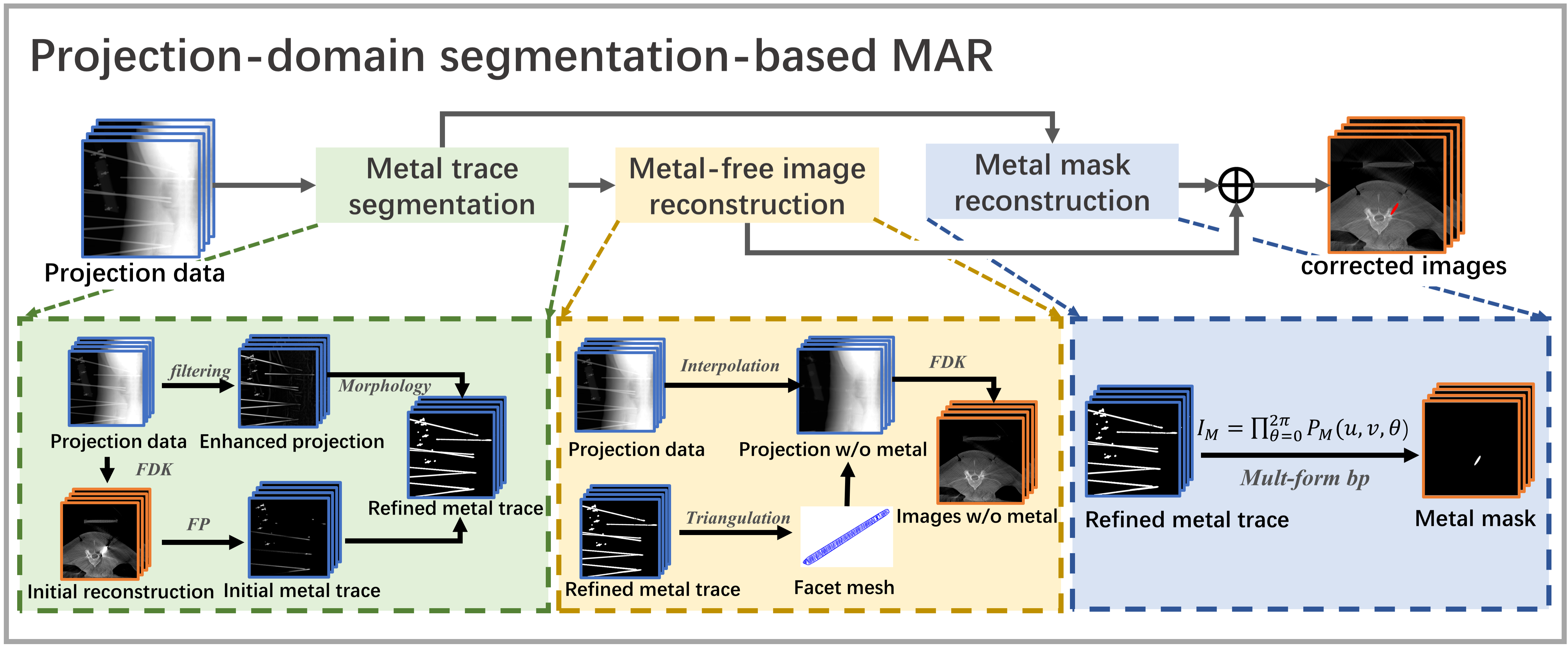}
    \caption{Flowchart of the proposed PDS-MAR algorithm.}
    \label{fig:flowchart}
\end{figure}
Fig. \ref{fig:flowchart} is a flowchart of the phases in the proposed algorithm. Reconstructed directly from the original rawdata is an uncorrected image. By thresholding the uncorrected image, the mask of metal is obtained. Simultaneously, tubular enhancement filtering is applied to enhance guidewires within projection images. Metal traces are segmented with morphological operations on enhanced projection views and projected metal masks. To inpaint metal traces in each projection view, a triangulation-based interpolation is then applied. Inpainted projection data and metal traces are used to reconstruct metal-free images and refined metal masks, respectively. The specific procedures are described in the sections that follow. 
\subsection{Metal trace segmentation}
This study focuses primarily on the segmentation of metallic guidewires. Guidewires are tubular objects in the projection domain, so techniques for enhancing vascular objects can be applied. Here we refer to the neurite enhancement filtering method proposed in \cite{Meijering2004} to enhance guidewires. This approach relies on a modified Hessian matrix $H'(x)$ defined with the following equation
\begin{equation}
    H'_{P*G_{\sigma}}(x)=H_{P*G{\sigma}}(x)+\alpha R^T_{\pi/2}H_{P*G{\sigma}}(x)R_{\pi/2}
\end{equation}
where $P*G_{\sigma}$ denotes the result of the input projection view $P$ Gaussian-smoothed with $\sigma$ variance, $H_{f*G{\sigma}}$ represents the Hessian matrix of the smoothed image, $R_{\pi/2}$ stands for the rotation matrix with angle $\pi/2$. Let $\lambda^{\sigma}_1(x)$ and $\lambda^{\sigma}_2(x)$ be the two eigenvalues of the Hessian matrix $H'_{P*G_{\sigma}}(x)$ and we have $|\lambda^{\sigma}_1(x)|>|\lambda^{\sigma}_2(x)|$, the output vesselness is then defined by
\begin{equation}
    V_{\sigma}(x)=
    \begin{cases}
        \lambda^{\sigma}_{1}(x) / \lambda^{\sigma}_{max}, &  \lambda^{\sigma}_{max} > 0\\
        0, & \lambda^{\sigma}_{max} \leq 0
    \end{cases}
    \label{eq:ves}
\end{equation}
where $\lambda^{\sigma}_{max}$ refers to the largest $\lambda^{\sigma}_1(x)$ among all pixels in the image. Eq. \ref{eq:ves} normalizes vesselness image to the range of $[0,1]$. To better model guidewires with different cross-sectional radii, vesselnesses under a set of different Gaussian variances $\Sigma={\sigma_i}, i=1,2,...,N$ are calculated, and the enhancement result is set as the maximal vesselness value on each pixel. 
\begin{equation}
    T(x)=max_{\sigma \in \Sigma}V_{\sigma}(x)
\end{equation}

Even though the tubular enhancement filtering step substantially enhances guidewire-affected regions, it also enhances other image boundaries, such as bone and body boundaries. A simple threshold $Th_{enh}$ will include a lot of undesired regions. Fortunately, the metal traces from the image domain segmentation are still available for guidance. The image domain metals are segmented with a threshold $Th_{metal}$ and then forward-projected into the image domain. Points with values greater than 0 are considered seed points for region-growing, and the resultant binary masks are considered metal-affected region masks. 

\subsection{Metal-free image  reconstruction}
Since guidewires are typically inserted horizontally, conventional 1-D linear interpolation in rows is likely to lose a great deal of tissue information. To enhance inpainting precision, we use a Delaunay triangulation-based 2-D interpolation method. 

To reduce computation costs, the input binary mask is first subdivided into connected components. The outer boundary of each connected component is extracted by subtracting itself from the morphologically dilated mask. The area within the outer boundary is then triangulated using the Delauney method \cite{delaunay1934sphere}. For each point $x$ inside a triangle facet $\Delta_{x_1x_2x_3}$, the interpolated value $P(x)$ is defined as a linear weighting of value on the vertices. 
\begin{equation}
    P(x)=w_{x_1}P(x_1)+w_{x_2}P(x_2)+w_{x_3}P(x_3)
\end{equation}
where the weights are calculated with the following equations.
\begin{equation}
    \begin{cases}
        w_{x_1}=\frac{\Vec{x_3x}\times\Vec{x_3x_2}}{\Vec{x_3x_1}\times\Vec{x_3x_2}} \\
        w_{x_2}=\frac{\Vec{x_3x}\times\Vec{x_3x_1}}{\Vec{x_3x_2}\times\Vec{x_3x_1}} \\
        w_{x_3}=1-w_{x_1}-w_{x_2}
    \end{cases}
\end{equation}

With metal traces interpolated in the projection domain, metal-free images can be reconstructed with the widely-accepted FDK algorithm \cite{feldkamp1984practical}. 

\subsection{Metal mask reconstruction}
Despite that we have already reconstructed a set of metal-free images, all information regarding the locations of metallic guidewires has been lost. Here, we present a method to restore image-domain metal masks $I_M(x)$  from projection-domain metal traces $P_M(u,v,\theta)$ based on multiplicative-form backprojection. Assuming that we have the perfect metal traces in the projection domain, forward-projecting an image-domain metal voxel will always find a projection-domain point inside the metal traces, whereas projecting a voxel outside the metallic convex hull will find points outside the metal traces in at least some views. The metal mask can therefore be reconstructed using the following equation:
\begin{equation}
    I_M(x)=\prod_{\theta=0}^{2\pi}P_M(u(x,\theta),v(x,\theta),\theta)
    \label{eq:rec}
\end{equation}
where $u(x,\theta)$ and $v(x,\theta)$ are the detector coordinate at view angle $\theta$ corresponding to image point $x$. 

Considering false negative points in metal trace segmentation which may introduce a large number of false negatives in $I_M$, we replace $P_M$ in Eq. \ref{eq:rec} with a soft mask $P^{soft}_M$ defined as follows
\begin{equation}
    P^{soft}_M=(1-\gamma)P_M+\gamma(1-P_M)
\end{equation}
where $\gamma$ is a weighting parameter in $(0,1)$. We usually choose $\gamma$ values close to 1 (like 0.9). After soft-masking, metal points get large $I_M$ values even if affected by false negatives on limited views while non-metal points get values close to 0. Metal masks are therefore segmented with a threshold $Th_{mask}$. 

\section{Experiments and Results}
\label{sec:res}
\subsection{Experiment settings}
All algorithms are implemented and tested in MATLAB R2021 with some operators implemented in C++/CUDA and compiled with mex/mexcuda in MATLAB. As for parameter settings, we set $\alpha=1/3$, $\Sigma=\{1,3,5,7,9\}$, $Th_{enh}=0$, $\gamma=0.9$ and $Th_{mask}=0.5$ for all experiments, as the results are not sensitive to these parameters. Parameter $Th_{metal}$ was set to 3000HU by default, and its settings will be discussed in \ref{subsec:humanbody}. 

\subsection{Digital phantom study}
\begin{table}[tbh]
    \caption{Geometrical Parameters of the Simulated CBCT Scan.}
    \centering
    \begin{tabular}{ c c c c}
        \Xhline{1px}
        Parameter & Value & Parameter & Value \\
        \hline
        DSD & 1140 mm & DSO & 617 mm \\
        detector pixels & \(1024\times{1024}\) & detector pixel size & \(0.45\times{0.45mm^2}\) \\
        volume size & \(512\times{512}\times{512}\) & voxel size & \(0.55\times{0.55}\times{0.55mm^3}\) \\
        view number & 300 & scan angle & 180°\\
        \Xhline{1px}
    \end{tabular}
    \label{tab:geo}
\end{table}
To quantitatively evaluate the performance of PDS-MAR, we simulated CBCT scans under a clinical intraoperative C-arm CBCT geometry with metallic guidewires inserted. XCIST \cite{wu2022xcist}, an open-source X-ray CT simulation toolkit, is used to simulate the projection data using a cone beam geometry. The detailed geometrical parameters are presented in Table \ref{tab:geo}, where $DSD$ represents the distance between source and detector, $DSO$ stands for distance between source and rotation axis. We used the default 110kV spectrum provided in XCIST without any pre-filtration for simulations. For the digital phantom, the voxelized female chest phantom provided along with XCIST (\url{https://github.com/xcist/phantoms-voxelized}) is adopted for simulation, which is also a part of the XCAT phantom dataset by Segars et al. \cite{Segars2010}. 

To simulate CBCT metal artifact, an additional material mask is manually outlined from a set of intraoperative clinical CBCT images using guidewires and applied to the digital phantom as the distribution of material 'iron'. Additionally, we have modified the phantom position offset to simulate CBCT data truncation during spine procedures. 

As the performance of MAR algorithms mainly depends on the projection-domain interpolation, here we first evaluate the projection-domain accuracy. Classical MAR methods (MAR-LI \cite{kalender1987reduction} and NMAR \cite{Meyer2010}) are also compared here to show the performance of the proposed algorithm. To evaluate the effectiveness of each component, ablation analysis with two different configurations is also included here: 1) MAR-tri, MAR with image-domain metal trace segmentation and triangulation-based interpolation; 2) PDS-MAR-LI, MAR with projection-domain metal trace segmentation and linear interpolation. 

\begin{figure}
    \centering
    \includegraphics[width=\textwidth]{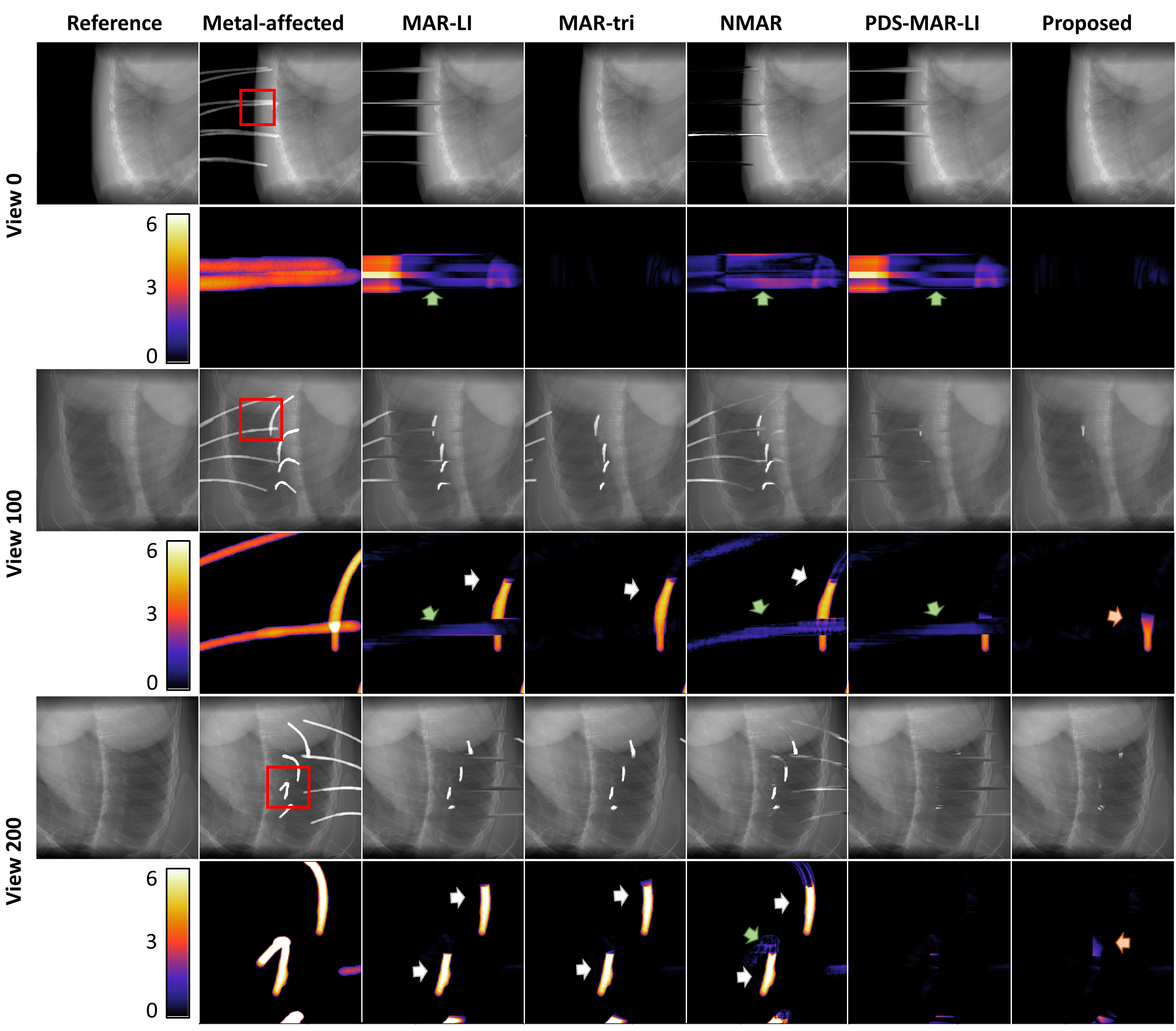}
    \caption{Projection domain interpolation results on 3 different views, with the enlarged absolute difference images between results and reference image corresponding to the red box in the following rows. From top to bottom are the projection data at view 0, view 100 and view 200, respectively. From left to right are the reference data simulated w/o metal, metal-affected data and results of MAR-LI,  MAR-tri, NMAR, PDS-MAR-LI, the proposed algorithm, respectively. }
    \label{fig:pres}
\end{figure}

Fig. \ref{fig:pres} provides a visual comparison of different algorithms, and results on 3 different views are depicted. The reference projection views are simulated using the digital phantom without metal implants under the same geometry. Compared to other algorithms, the proposed algorithm generate corrected projection views closest to the reference views. To better view the details, Fig. \ref{fig:pres} also shows the enlarged projection-domain blocks corresponding to the red rectangles. The errors of MAR algorithms mainly come from two aspects, false segmentation and non-ideal interpolation. MAR-LI, MAR-tri and NMAR adopt image-domain segmentation, which introduces obvious false segmentations (white arrows in Fig. \ref{fig:pres}). Despite the fact that metals can be easily segmented with a threshold on reconstructed phantom images, there is no way for image-domain methods to get metal traces outside the reconstruction FoV. Besides, 1-D linear interpolation produces over-smoothed rows (green arrows in Fig. \ref{fig:pres}). This characteristic is particularly harmful when metal implants are long in the horizontal direction (View 0 in Fig. \ref{fig:pres}), which is, however, common for guidewires. Triangulation-based interpolation, on the other hand, inpaints the void regions according to 2-D information and results in interpolated views much closer to references. There are still some defects shown in the results of the proposed algorithm (orange arrows in Fig. \ref{fig:pres}). Still, generally, the proposed algorithm produces the best-interpolated projection data among all algorithms. 

\begin{table}[tbh]
    \centering
    \caption{Quantitative evaluation of different algorithms in projection-domain}
    \begin{tabular}{c|c c c | c c c}
        \Xhline{1px}
        \multirow{2}{*}{Method} & \multicolumn{3}{c|}{Segmentation} & \multicolumn{3}{c}{Interpolation} \\
         & Precision$\uparrow$ & Recall$\uparrow$ & Dice$\uparrow$ & RMSE$\downarrow$ & PSNR(dB)$\uparrow$ & SSIM(\%)$\uparrow$ \\
         \hline
         MAR-LI & 0.8720 & 0.8071 & 0.8383 & 0.5267 & 22.27 & 93.77 \\
         MAR-tri & 0.8720 & 0.8071 & 0.8383 & 0.4149 & 25.34 & 97.11 \\
         NMAR & 0.8720 & 0.8071 & 0.8383 & 1.7211 & 16.58 & 92.39 \\
         PDS-MAR-LI & \textbf{0.9092} & \textbf{0.9470} & \textbf{0.9277} & 0.2109 & 32.88 & 94.95 \\
         \hline
         PDS-MAR & \textbf{0.9092} & \textbf{0.9470} & \textbf{0.9277} & \textbf{0.0514} & \textbf{45.09} & \textbf{98.41} \\
        \Xhline{1px}
    \end{tabular}
    \label{tab:quan_proj}
\end{table}
To quantitatively evaluate the performance of the proposed algorithm, several metrics are calculated between the outputs and the references regarding both segmentation accuracy and final interpolation quality. For segmentation, we compute Precision, Recall, and Dice index between the generated metal trace and the simulated projection data with the metal object only. Commonly used metrics including Root Mean Square Error (RMSE), Peak-Signal-to-Noise Ratio (PSNR), and Structural Similarity Index Metric (SSIM) are also calculated between interpolated projection data and reference data to compare interpolation similarity. The results are displayed in Table \ref{tab:quan_proj}. The proposed projection-domain segmentation method achieves a dice index of 0.9188, whereas conventional image-domain segmentation only gets 0.8141. The projection-domain method achieves a $\sim4\%$ improvement in segmentation precision and a $\sim14\%$ improvement in recall over the image-domain method. The significant increase in recall is primarily attributable to the successful segmentation of metal traces outside the reconstruction field of view. The proposed method also obtains the highest metric values for final interpolation similarity compared to all other comparing methods. From the results, we get the following observations. (1) PDS-MAR-LI/PDS-MAR achieves better interpolation than MAR-LI/MAR-tri. This proves that projection-domain segmentation is more appropriate than image-domain segmentation. (2) Triangulation-based interpolation methods (MAR-tri/PDS-MAR) achieve better results than linear interpolation methods (MAR-LI/PDS-MAR-LI). (3) NMAR gets the worst results on all metrics. This may result from poor tissue/bone segmentation in the image domain. 

\begin{figure}
    \centering
    \includegraphics[width=0.9\textwidth]{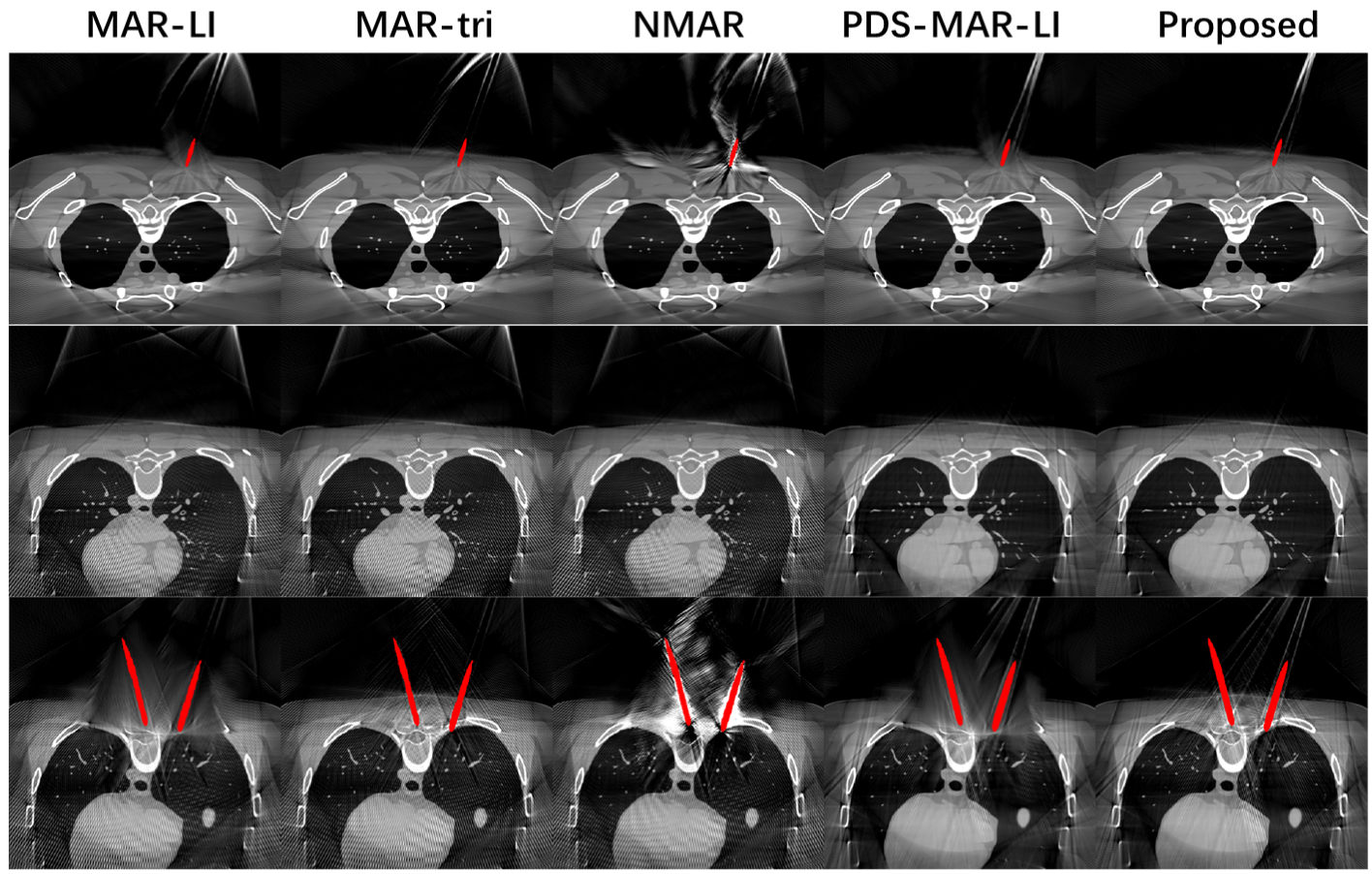}
    \caption{Reconstruction results of different MAR methods. Metal masks are labeled in red. All images are displayed under the window of (C=-250HU, W=1500HU). }
    \label{fig:pha_img}
\end{figure}
\begin{table}[tbh]
    \centering
    \caption{Quantitative evaluation of different algorithms in image-domain}
    \begin{tabular}{c|c c c | c c c}
        \Xhline{1px}
        \multirow{2}{*}{Method} & \multicolumn{3}{c|}{Metal Segmentation} & \multicolumn{3}{c}{Artifact Reduction} \\
         & Precision$\uparrow$ & Recall$\uparrow$ & Dice$\uparrow$ & RMSE(HU)$\downarrow$ & PSNR(dB)$\uparrow$ & SSIM(\%)$\uparrow$ \\
         \hline
         MAR-LI & 0.7680 & 0.9602 & 0.8534 & 109.10 & 34.28 & 98.03 \\
         MAR-tri & 0.7680 & 0.9602 & 0.8534 & 102.93 & 35.99 & 98.65 \\
         NMAR & 0.7680 & 0.9602 & 0.8534 & 600.90 & 26.37 & 95.29 \\
         ID-ideal & - & - & - & 91.82 & 39.30 & 98.28 \\
         PDS-MAR-LI & \textbf{0.7746} & \textbf{0.9912} & \textbf{0.8696} & 54.71 & 38.73 & 99.18 \\
         \hline
         PDS-MAR & \textbf{0.7746} & \textbf{0.9912} & \textbf{0.8696} & \textbf{41.24} & \textbf{41.32} & \textbf{99.63} \\
        \Xhline{1px}
    \end{tabular}
    \label{tab:quan_img}
\end{table}
Fig. \ref{fig:pha_img} gives a visual comparison between the reconstruction results of different algorithms. Generally, image domain segmentation-based methods (MAR-LI, MAR-tri and NMAR) suffer from severe streak artifacts resulting from metal objects out of scanning FoV. Linear interpolation-based algorithms (MAR-LI, PDS-MAR-LI) suffer from over-smoothing near metal objects, and tissue contrast details get lost. The proposed algorithm shows much-reduced streak artifacts and sharper tissue boundaries around metal implants than other algorithms. However, the proposed algorithm does not eliminate streak artifacts totally, since there are still interpolation failures in the projection domain (orange arrows in Fig. \ref{fig:pres}). 

Quantitative evaluation results on the image domain are provided in Table \ref{tab:quan_img}. We first compared the metal mask accuracy between image domain thresholding and the proposed metal mask reconstruction method. The image domain thresholding method achieves a pretty good dice score in the phantom study. However, the success in thresholding mainly results from low image noise, low bone value and much-suppressed image artifacts in our simulated data, and cannot be reproduced on authentic CBCT images (see Fig. \ref{fig:met_examp}). The proposed algorithm achieves a dice score $\sim1\%$ higher than image domain thresholding does. As for artifact reduction, we compute RMSE, PSNR and SSIM metrics between MAR results and reference images from metal-free data. The proposed algorithm shows significant improvements in all metrics compared to other algorithms. 

\subsection{Ideal image-domain method study}
\label{sec:ideal}
According to the results of MAR-tri at view 0 in Fig. \ref{fig:pres}, metal traces are nearly eliminated at this view. The near-perfect interpolation results from a near-perfect segmentation at this view, indicating that metal objects within the scanning FoV in the image domain are nearly accurately segmented. However, the results of the reconstruction indicate that metal artifacts cannot be eliminated under these conditions. There are metal objects outside the scanning FoV that can be observed from other perspectives, and the artifacts introduced by these metal objects cannot be corrected using image-domain segmentation. 

\begin{figure}
    \centering
    \includegraphics[width=0.8\textwidth]{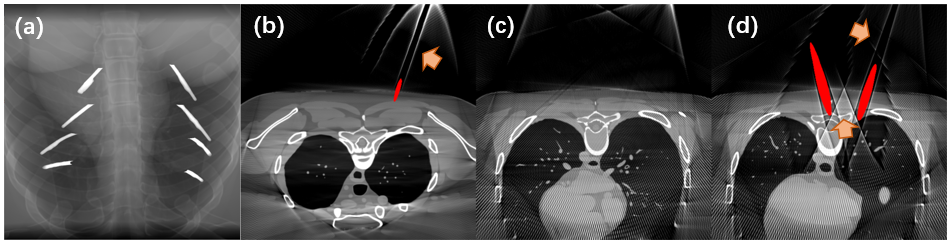}
    \caption{Projection- and image-domain results with an ideal image-domain segmentation-based method. (a) projection result at view 150; (b),(c),(d) 3 reconstructed slices. }
    \label{fig:ideal}
\end{figure}

To prove this conjecture, we conducted an experiment on an ideal image-domain segmentation-based MAR method (ID-ideal). In ID-ideal, the image-domain metal mask is generated by masking the ground truth with scanning FoV. The metal mask is then forward-projected into the projection to produce metal traces. For interpolation, we directly replace the values within the metal traces with corresponding values in the reference projection data (simulated without metal), which should be considered perfect interpolation. The resulting projection data at a view and reconstruction outputs on 3 slices are presented in \ref{fig:ideal}. From the results, we see that metal traces are not completely corrected in the projection domain, which further introduces severe streak artifacts to the image domain. There are also streak artifacts introduced by boundary precision issues (orange arrows in Fig. \ref{fig:ideal} (b,d)). Quantitative evaluations of the image domain are also provided in Table \ref{tab:quan_img}. The metrics attained by PDS-MAR are superior to those attained by ID-ideal. 

\subsection{Animal body study}
\label{subsec:anibody}
\begin{comment}
\begin{figure}
    \centering
    \includegraphics[width=0.9\textwidth]{imgs/sheep_proj.png}
    \caption{Three metal-corrected projection views for the sheep body data. From left to right are the original data without MAR and results of MAR-LI, MAR-tri, NMAR, PDS-MAR-LI and the proposed PDS-MAR method, respectively. }
    \label{fig:sheep_proj}
\end{figure}
\end{comment}
\begin{figure}
    \centering
    \includegraphics[width=0.9\textwidth]{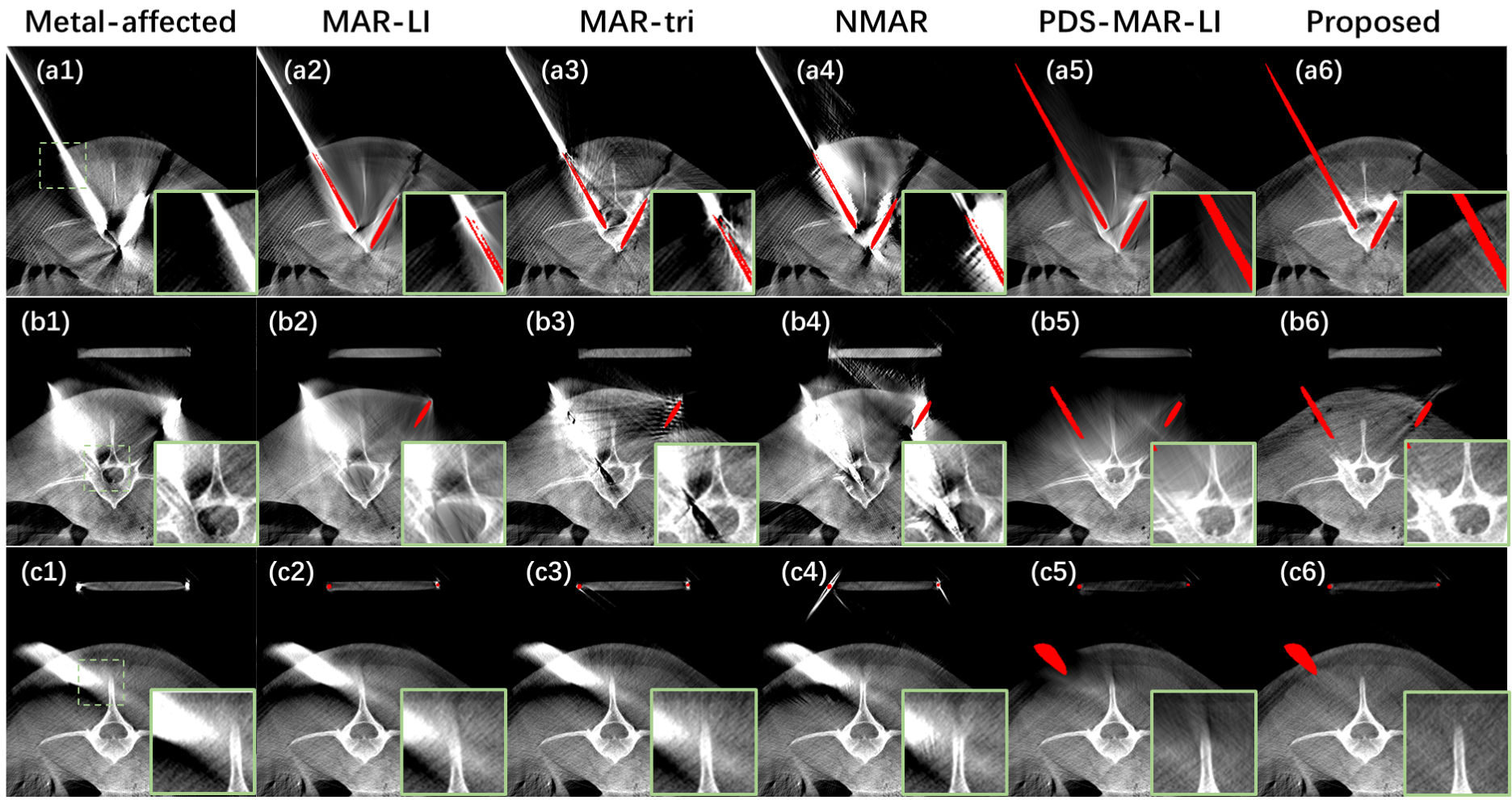}
    \caption{Three reconstructed slices for sheep body data. From left to right are the reconstruction without MAR and results of MAR-LI, MAR-tri, NMAR, PDS-MAR-LI and the proposed PDS-MAR method, respectively. Corresponding metal masks are labeled in red. All images are displayed under the window of (C=250HU, W=1500HU). }
    \label{fig:sheep_img}
\end{figure}
The proposed algorithm is evaluated on a set of animal body data. The back of a deceased sheep was scanned with Kirschner wires (K-wires) inserted by an expert. These projection data were acquired using a commercial intraoperative CBCT system (B51s) from First-Imaging Medical Equipment Co., Ltd. All original data were initially pre-processed with commercial data correction software from First-Imaging (including field correction, denoising, and water beam hardening correction). 

\begin{comment}
    Fig. \ref{fig:sheep_proj} displays the metal-interpolated projection data, while Fig. \ref{fig:sheep_img} provides the reconstructed results. Unlike in digital phantom data, results on animal body data suffer from severe segmentation failures in the image domain with a threshold. Some K-wires are partially segmented (the left one in Fig. \ref{fig:sheep_img} (a1-a4)) some are not found at all (the left one in Fig. \ref{fig:sheep_img} (b1-b4), the one in Fig. \ref{fig:sheep_img} (c1-c4)). The missegmentation in the image domain causes a significant metal trace missing in the projection domain (red rectangles in Fig. \ref{fig:sheep_proj}) which further leads to in uncorrected metal artifacts in the reconstructed images. The proposed projection domain segmentation method also fails in some projection views (green rectangles in Fig. \ref{fig:sheep_proj} (b5-b6)), but segmentation failures on a few views have little impact on the reconstruction results. Also, the projection domain segmentation method cannot segment isolated metal objects totally outside the reconstruction FoV (green rectangles in Fig. \ref{fig:sheep_proj} (c5-c6)) because it still relies on image domain segmentation results as seed points. Generally, the proposed method reconstructs image slices with substantially reduced metal artifacts and tissue details that are preserved. In addition, the proposed metal mask reconstruction method is able to preserve the correct image domain metal mask even when the thresholding strategy fails. 
\end{comment}
Fig. \ref{fig:sheep_img} provides the reconstructed results from sheep body data. Unlike in digital phantom data, results on animal body data suffer from severe segmentation failures in the image domain with a threshold. Some K-wires are partially segmented (the left one in Fig. \ref{fig:sheep_img} (a1-a4)) some are not found at all (the left one in Fig. \ref{fig:sheep_img} (b1-b4), the one in Fig. \ref{fig:sheep_img} (c1-c4)). The missegmentation in the image domain causes a significant metal trace missing in the projection domain which further leads to uncorrected metal artifacts in the reconstructed images. Generally, the proposed algorithm reconstructs image slices with substantially reduced metal artifacts and tissue details that are preserved. In addition, the proposed metal mask reconstruction method is able to preserve the correct image domain metal mask even when the thresholding strategy fails. 

\begin{figure}
    \centering
    \includegraphics[width=0.5\textwidth]{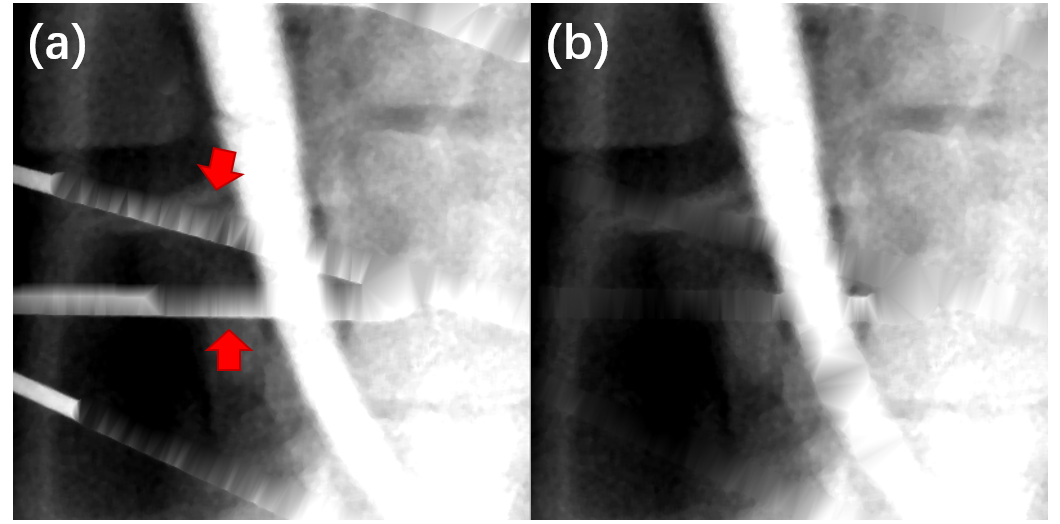}
    \caption{A projection domain patch example of boundary inaccuracy causing saw-toothed inpainting for triangularized interpolation. (a). inpainting result of MAR-tri; (b) inpainting result of the proposed method. }
    \label{fig:bound}
\end{figure}

Some K-wires are nearly correctly segmented in the image domain (the right one in Fig. \ref{fig:sheep_img} (b1-b4)), but still get slight deviations from the real metal traces in the projection domain after the forward projection. These deviations sometimes come from the geometrical randomness of the entire CBCT system. Triangulation-based interpolation is ineffective in this situation and results in saw-toothed inpainting results (red arrows in Fig. \ref{fig:bound} (a)). The saw-toothed projection data further introduces severe streak artifacts to the reconstructed images as a result of ramp filtering (Fig. \ref{fig:sheep_img} (b3) as an example). As segmentation in the projection domain is more precise than in the image domain, the proposed method generates interpolation output that is substantially smoother. 

\subsection{Human body phantom study}
\label{subsec:humanbody}
\begin{figure}
    \centering
    \includegraphics[width=0.9\textwidth]{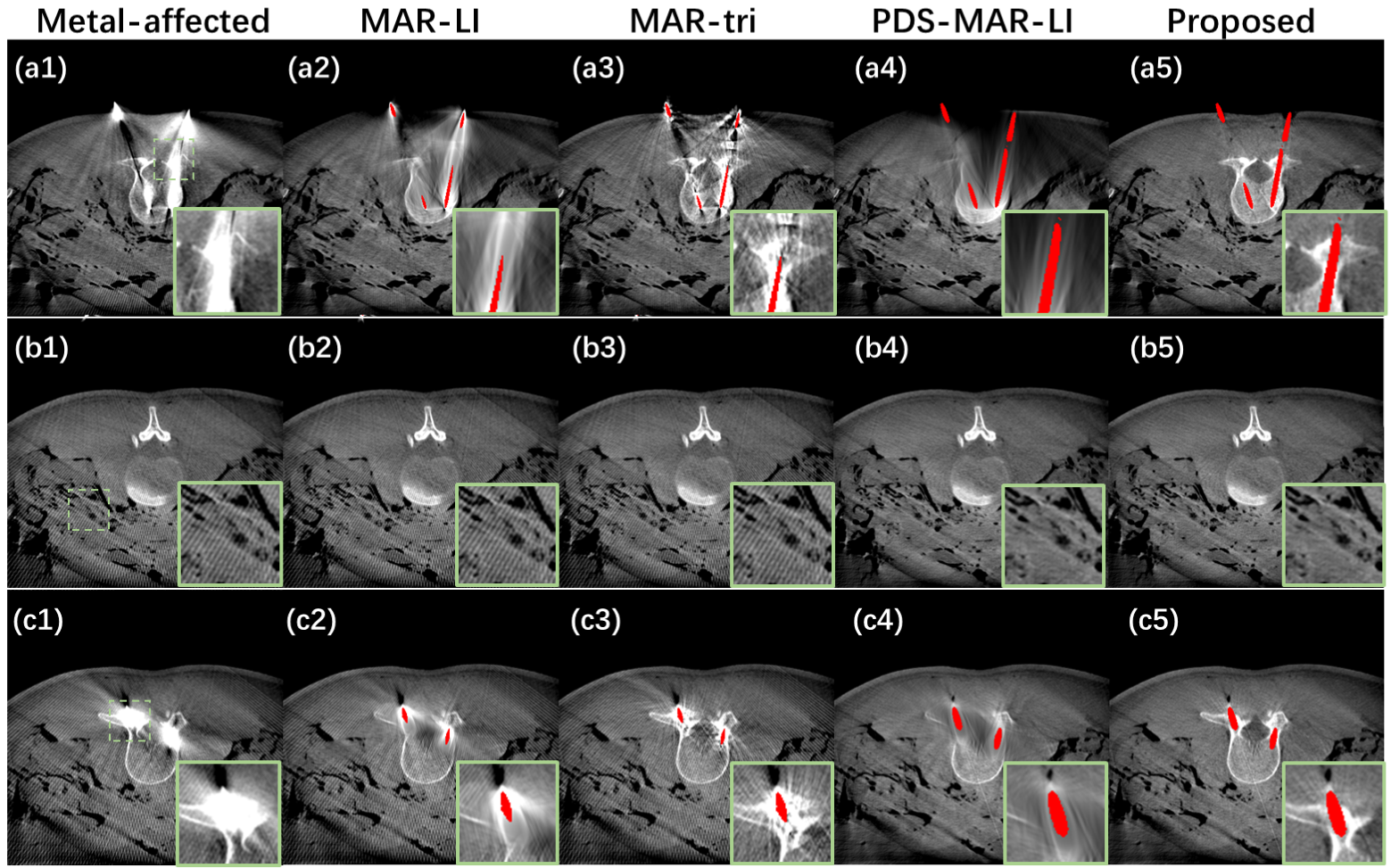}
    \caption{Three reconstructed slices for the human abdominal phantom. From left to right are the reconstruction without MAR and results of MAR-LI, MAR-tri, PDS-MAR-LI and the proposed PDS-MAR method, respectively. Corresponding metal masks are labeled in red. All images are displayed under the window of (C=250HU, W=1500HU). }
    \label{fig:abdomen}
\end{figure}

Two human-body phantoms (an abdominal phantom and a head phantom) are scanned with guidewires inserted for educational purposes. The projection data used in this section were also collected with a commercial intraoperative CBCT system from First-Imaging Medical Equipment Co., Ltd. All original data were pre-processed as described in \ref{subsec:anibody}. 

The reconstruction results for the abdominal phantom are depicted in Fig. \ref{fig:abdomen}. Similar to the animal body results, conventional image domain segmentation suffers from segmentation faults and results in incomplete metal artifact suppression (see Fig. \ref{fig:abdomen} (c2-c3)) and secondary artifacts (see Fig. \ref{fig:abdomen} (a3)). The proposed algorithm shows superior ability in metal artifact reduction, and the metal masks are well-reconstructed with our multiplicative-form backprojection method. Furthermore, streak artifacts introduced by metal objects outside FoV are still observed even on slices without metal in Fov for image domain segmentation methods (bottom of Fig. \ref{fig:abdomen} (b2-b3)) while being totally eliminated by the proposed method (Fig. \ref{fig:abdomen} (b5)).

\begin{figure}
    \centering
    \includegraphics[width=0.9\textwidth]{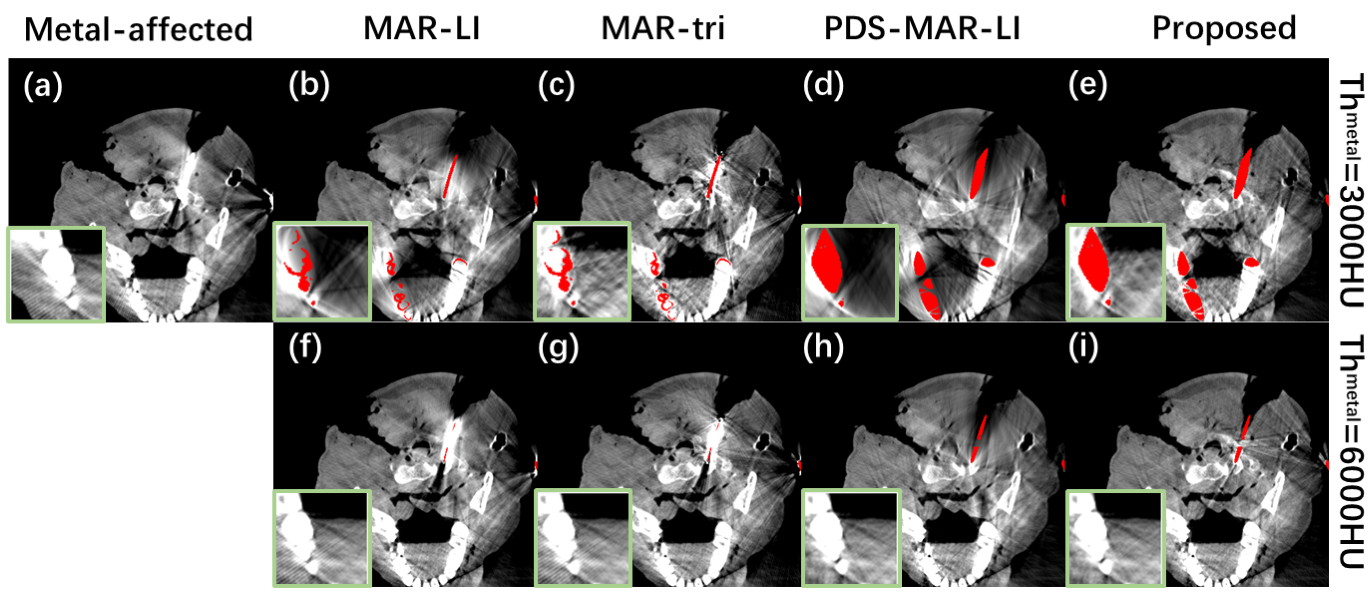}
    \caption{A reconstructed axial slice on the human head phantom. From left to right are the reconstruction without MAR and results of MAR-LI, MAR-tri, PDS-MAR-LI and the proposed PDS-MAR algorithm, respectively. Corresponding metal masks are labeled in red. All images are displayed under the window of (C=250HU, W=1500HU). }
    \label{fig:head}
\end{figure}

\begin{figure}
    \centering
    \includegraphics[width=0.9\textwidth]{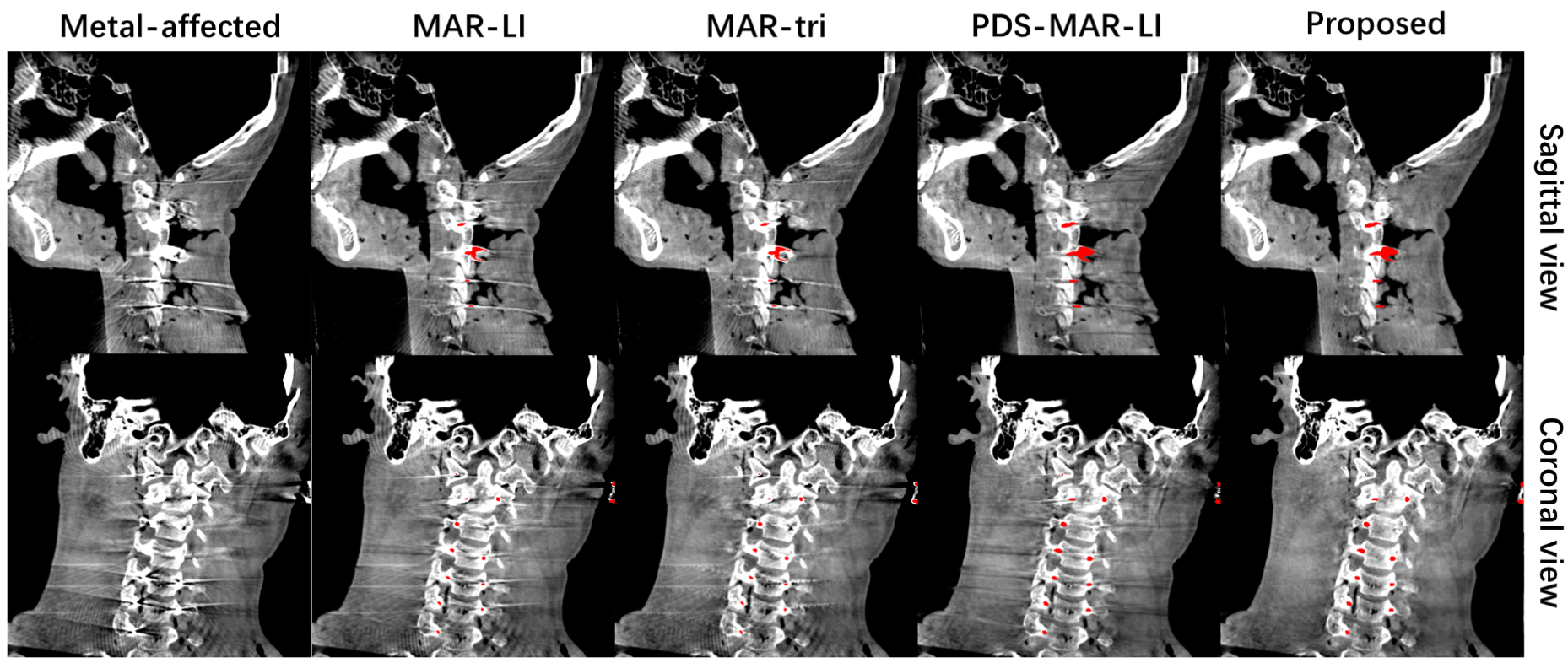}
    \caption{Results for the human head phantom on sagittal and coronal views. The metal threshold was set to 6000HU for all images. Corresponding metal masks are labeled in red. All images are displayed under the window of (C=250HU, W=1500HU). }
    \label{fig:sag_cor}
\end{figure}

The reconstruction results for the head phantom are depicted in Fig. \ref{fig:head}. These results reveal another shortcoming of image domain segmentation, namely the difficulty in threshold choice. Referring to Fig. \ref{fig:head} as an example, we compared results under $Th_{metal}=3000HU$ (b-e) and $Th_{metal}=6000HU$ (f-i) for different methods. Under a lower threshold, image domain methods can segment metal better. However, tooth regions are identified as metal objects due to their high attenuation properties, which introduce secondary (see Fig. \ref{fig:head} (b,c)). Under the higher threshold, tooth regions are not included; however, guidewires are not recognized and metal artifacts are not corrected. The proposed also suffers from segmentation faults on teeth under the threshold of 3000HU but performs well under the threshold of 6000HU. A slice in sagittal view and a slice in coronal view are displayed in Fig. \ref{fig:sag_cor}. Besides guidewires, there are also metal screws inserted in the head phantom and the proposed algorithm can effectively deal with metal artifacts introduced by these screws. 

\section{Discussion}
\label{sec:dis}
Metal artifact is a long-standing problem for intraoperative CBCT imaging in MISS. In section \ref{sec:ideal}, we experimentally proved that the image-domain segmentation-based MAR methods can never eliminate metal artifacts when there are metal objects outside scanning FoV. This situation, however, is prevalent in CBCT-based intraoperative imaging. Limited by the size of the flat panel detector, CBCTs are incapable of achieving a large scanning FOV like most MSCTs. Also, objects like guidewires can easily get outside the FoV due to their lengths. Therefore, projection-domain segmentation of metallic objects is urgently required. 
\begin{figure}
    \centering
    \includegraphics[width=0.7\textwidth]{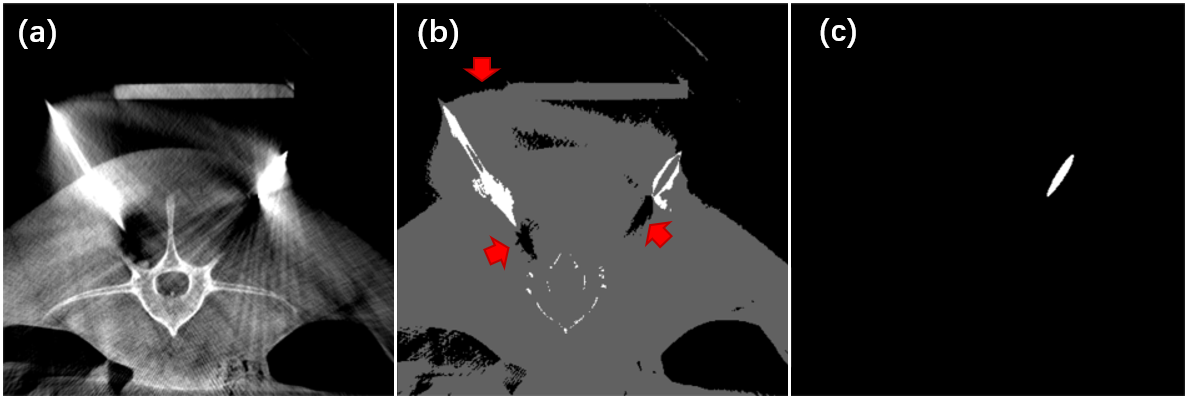}
    \caption{An example slice on animal body data for NMAR algorithm. (a) Initial reconstruction; (b) Prior image by segmentation; (c) Metal mask. }
    \label{fig:nmar}
\end{figure}

CBCT's inferior image quality is another cause for the failure of conventional MAR techniques. Scattering, noise, limited scan angle, and other factors introduce artifacts. On our CBCT dataset, the widely accepted NMAR algorithm for metal artifact reduction in CT images performs even worse than MAR-LI. This phenomenon is a result of NMAR's reliance on image domain segmentation outcomes. Using Fig. \ref{fig:nmar} as an illustration, NMAR relies on image domain segmentation to get a prior image. Due to severe image domain artifacts (red arrows in Fig. \ref{fig:nmar} (b)),  the prior image is improperly generated, resulting in severe interpolation errors in the projection domain. 

\begin{figure}
    \centering
    \includegraphics[width=0.7\textwidth]{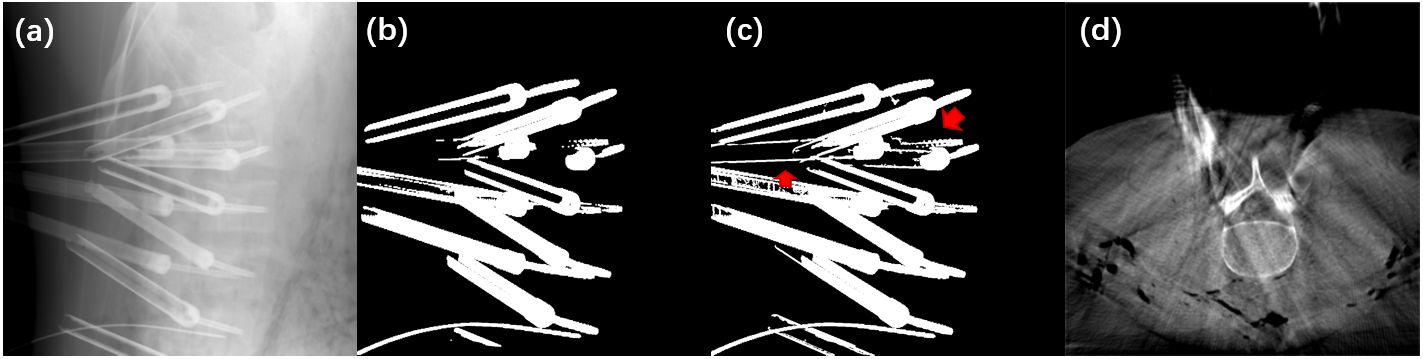}
    \caption{Results on an abdominal phantom scan with Ti screws inserted. (a) Projection data; (b) Metal trace from image domain segmentation; (c) Metal trace from the proposed algorithm; (d) A reconstructed slice (metal mask not applied). }
    \label{fig:screw}
\end{figure}

Segmenting metal traces directly on the projection domain ought to be more accurate than the image-domain-segmentation + forward-projection way. However, segmentation in the projection domain is never easy due to the variation in both metal shapes and values. This paper focuses on a specific mission to segment metallic guidewires in MISS, which converts the metal into a tubular shape. In this endeavor, tubular enhancement filtering demonstrates exceptional performance, and metal traces are successfully extracted by combining information from the image domain. Experiments indicate that the proposed method is also effective with metallic objects such as small iron balls and stainless steel screws. Nonetheless, additional experiments demonstrate that the proposed procedure is incapable of handling Ti screws with handles. Though the proposed method generates better metal traces compared to image domain methods (Fig. \ref{fig:screw} (b,c)), there are still missing portions at the screw head and the handles (red arrows in Fig. \ref{fig:screw} (c)), which results in artifacts in the image domain (Fig. \ref{fig:screw} (d)). How to cope with metal artifacts introduced by Ti screws is still an open issue, and we will be focusing our future research on this issue. 

Triangulation-based interpolation has been proven effective in multislice helical CT \cite{yu2009metal, Abdoli2011}, but has hardly been tested on CBCT datasets. In this work, we show that the Delaunay triangulation-based interpolation is a powerful tool in CBCT metal artifact reduction but it depends heavily on the metal trace accuracy. This method may introduce severe secondary artifacts with biased metal traces and result in even worse image quality compared to 1-D linear interpolation. The proposed projection domain segmentation method is an excellent complement to triangulation-based interpolation for providing more precise metal traces. 

\begin{figure}
    \centering
    \includegraphics[width=0.8\textwidth]{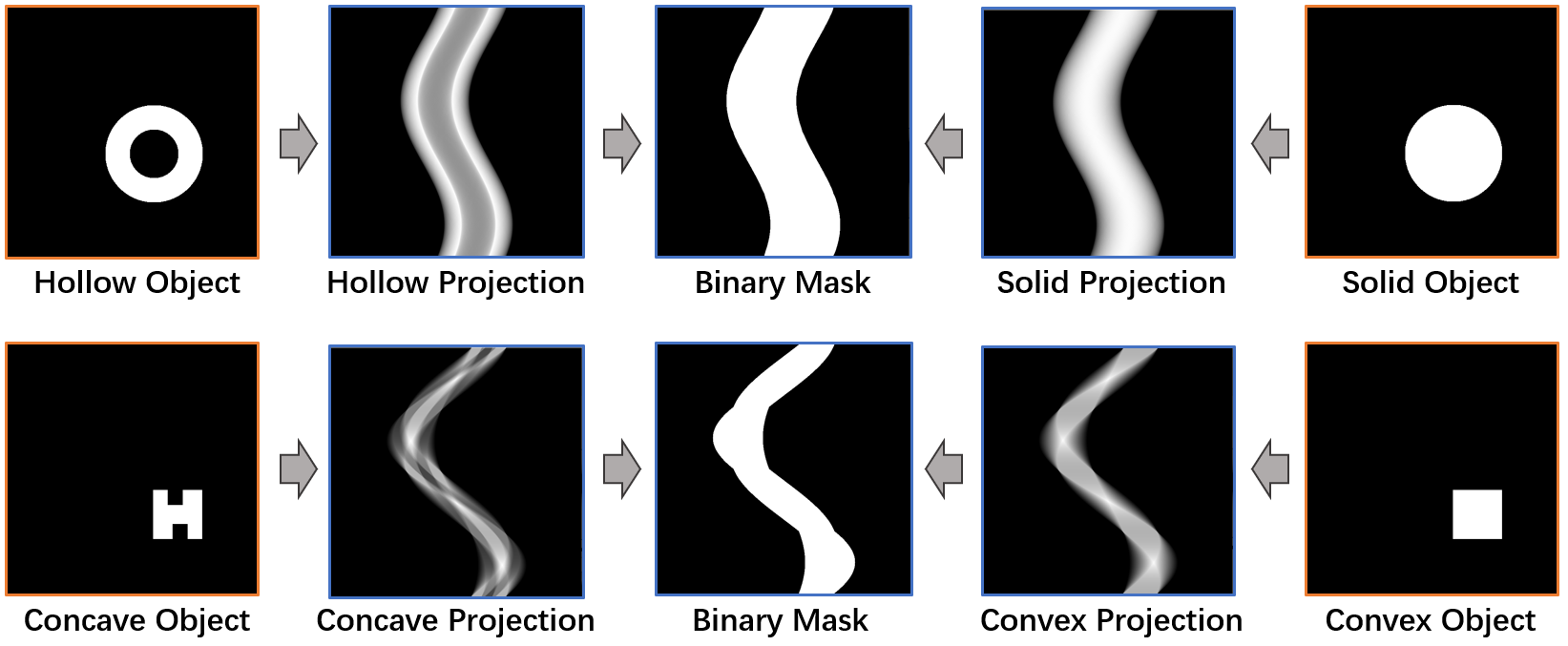}
    \caption{Illustration of the binary mask reconstruction problem in projection domain segmentation-based MAR algorithms. The hollow object and its corresponding solid object share the same projection domain binary mask. The concave object and its minimal convex hull also share the same projection domain binary mask. }
    \label{fig:hol_con}
\end{figure}

The retention of precise metal masks in the image domain is another problem for MAR algorithms based on projection domain segmentation. In this paper, we propose a multiplicative-form backprojection-based metal mask reconstruction method that performs well on CBCT images with guidewires. Nonetheless, it does not completely resolve the issue. The projection domain masks lose information on the thicknesses of metal materials, making it impossible to retain hollow or concave metal masks. Two 2-D examples are presented in Fig. \ref{fig:hol_con}. In the projection domain, the solid circle and hollow ring have distinct characteristics, but when converted to binary masks, they yield the exact same result, as do the concave object and its minimal convex hull. In CBCT, the information in the z-direction weakens this problem (screws in Fig. \ref{fig:sag_cor}), but it still presents on some slices (Fig. \ref{fig:head} (a5)). We are going to work in the future to incorporate metal thickness information into projection domain masks, which is the only way to solve this issue. 

\section{Conclusion}
\label{sec:con}
In this work, we prove that conventional image-domain segmentation-based MAR algorithms cannot eliminate metal artifacts for intraoperative CBCT images with guidewires, and present PDS-MAR to solve this problem. We employ tubular enhancement filtering-based metal trace segmentation and Delaunay triangulation-based interpolation in PDS-MAR. Results on both simulation and real CBCT datasets show the extraordinary artifact suppression performance of our algorithm. Moreover, a novel multiplicative-form backprojection-based method is also proposed here in PDS-MAR to retain image domain metal masks from metal traces. The concept of projection-domain metal segmentation would advance MAR techniques in CBCT and has the potential to push forward the use of intraoperative CBCT in human-handed and robotic-assisted MISS. 

\section*{Acknowledgements}
This work was supported in part by the National Natural Science Foundation of China under Grant T2225025, in part by the National Key Research and Development Program of China under Grant 2022YFC2408500, in part by the Key Research and Development Programs in Jiangsu Province of China under Grant BE2021703 and BE2022768, and also in part by the Key Research and Development Program of Zhejiang Province under Grant 2021C03029.

\section*{References}
\bibliographystyle{unsrt}
\bibliography{refs}

\begin{thebibliography}{10}

\bibitem{yoon2019evolution}
Jang~W Yoon and Michael~Y Wang.
\newblock The evolution of minimally invasive spine surgery: Jnspg 75th
  anniversary invited review article.
\newblock {\em Journal of Neurosurgery: Spine}, 30(2):149--158, 2019.

\bibitem{goldstein2016perioperative}
Christina~L Goldstein, Kevin Macwan, Kala Sundararajan, and Y~Raja Rampersaud.
\newblock Perioperative outcomes and adverse events of minimally invasive
  versus open posterior lumbar fusion: meta-analysis and systematic review.
\newblock {\em Journal of Neurosurgery: Spine}, 24(3):416--427, 2016.

\bibitem{wang2017systematic}
Xuan Wang, Benny Borgman, Simona Vertuani, and Jonas Nilsson.
\newblock A systematic literature review of time to return to work and narcotic
  use after lumbar spinal fusion using minimal invasive and open surgery
  techniques.
\newblock {\em BMC Health Services Research}, 17(1):1--14, 2017.

\bibitem{spetzger2013past}
Uwe Spetzger, Andrej~Von Schilling, Gerd Winkler, J{\"u}rgen Wahrburg, and
  Alexander K{\"o}nig.
\newblock The past, present and future of minimally invasive spine surgery: a
  review and speculative outlook.
\newblock {\em Minimally Invasive Therapy \& Allied Technologies},
  22(4):227--241, 2013.

\bibitem{vaishnav2019current}
Avani~S Vaishnav, Yahya~A Othman, Sohrab~S Virk, Catherine~Himo Gang, and
  Sheeraz~A Qureshi.
\newblock Current state of minimally invasive spine surgery.
\newblock {\em Journal of spine surgery}, 5(Suppl 1):S2, 2019.

\bibitem{park2020minimally}
Jiwon Park, Dae-Woong Ham, Byung-Taek Kwon, Sang-Min Park, Ho-Joong Kim, and
  Jin~S Yeom.
\newblock Minimally invasive spine surgery: techniques, technologies, and
  indications.
\newblock {\em Asian spine journal}, 14(5):694, 2020.

\bibitem{kim2008use}
Choll~W Kim, Yu-Po Lee, William Taylor, Ahmet Oygar, and Woo~Kyung Kim.
\newblock Use of navigation-assisted fluoroscopy to decrease radiation exposure
  during minimally invasive spine surgery.
\newblock {\em The Spine Journal}, 8(4):584--590, 2008.

\bibitem{tjardes2010image}
Thorsten Tjardes, Sven Shafizadeh, Dieter Rixen, Thomas Paffrath, Bertil
  Bouillon, Eva~S Steinhausen, and Holger Baethis.
\newblock Image-guided spine surgery: state of the art and future directions.
\newblock {\em European spine journal}, 19:25--45, 2010.

\bibitem{kraus2013image}
Michael Kraus, Sebastian Weckbach, Almut Jones, Gert Krischak, Florian Gebhard,
  and Hendrik Sch{\"o}ll.
\newblock Image guidance shortens the learning curve for k-wire placement--an
  experimental study.
\newblock {\em The International Journal of Medical Robotics and Computer
  Assisted Surgery}, 9(1):52--57, 2013.

\bibitem{overley2017navigation}
Samuel~C Overley, Samuel~K Cho, Ankit~I Mehta, and Paul~M Arnold.
\newblock Navigation and robotics in spinal surgery: where are we now?
\newblock {\em Neurosurgery}, 80(3S):S86--S99, 2017.

\bibitem{kochanski2019image}
Ryan~B Kochanski, Joseph~M Lombardi, Joseph~L Laratta, Ronald~A Lehman, and
  John~E O’Toole.
\newblock Image-guided navigation and robotics in spine surgery.
\newblock {\em Neurosurgery}, 84(6):1179--1189, 2019.

\bibitem{tonetti2020role}
J{\'e}r{\^o}me Tonetti, Mehdi Boudissa, Gael Kerschbaumer, and Olivier Seurat.
\newblock Role of 3d intraoperative imaging in orthopedic and trauma surgery.
\newblock {\em Orthopaedics \& Traumatology: Surgery \& Research},
  106(1):S19--S25, 2020.

\bibitem{tkatschenko2020navigated}
Dimitri Tkatschenko, Paul Kendlbacher, Marcus Czabanka, Georg Bohner, Peter
  Vajkoczy, and Nils Hecht.
\newblock Navigated percutaneous versus open pedicle screw implantation using
  intraoperative ct and robotic cone-beam ct imaging.
\newblock {\em European Spine Journal}, 29:803--812, 2020.

\bibitem{siewerdsen2020cone}
Jeffrey~H Siewerdsen.
\newblock Cone-beam ct systems.
\newblock {\em Computed Tomography: Approaches, Applications, and Operations},
  pages 11--26, 2020.

\bibitem{schafer2020technology}
Sebastian Schafer and Jeffrey~H Siewerdsen.
\newblock Technology and applications in interventional imaging: 2d x-ray
  radiography/fluoroscopy and 3d cone-beam ct.
\newblock In {\em Handbook of Medical Image Computing and Computer Assisted
  Intervention}, pages 625--671. Elsevier, 2020.

\bibitem{gjesteby2016metal}
Lars Gjesteby, Bruno De~Man, Yannan Jin, Harald Paganetti, Joost Verburg,
  Drosoula Giantsoudi, and Ge~Wang.
\newblock Metal artifact reduction in ct: where are we after four decades?
\newblock {\em Ieee Access}, 4:5826--5849, 2016.

\bibitem{kalender1987reduction}
Willi~A Kalender, Robert Hebel, and Johannes Ebersberger.
\newblock Reduction of ct artifacts caused by metallic implants.
\newblock {\em Radiology}, 164(2):576--577, 1987.

\bibitem{Meyer2010}
Esther Meyer, Rainer Raupach, Michael Lell, Bernhard Schmidt, and Marc
  Kachelrie{\ss}.
\newblock {Normalized metal artifact reduction (NMAR) in computed tomography}.
\newblock {\em Medical Physics}, 37(10):5482--5493, oct 2010.

\bibitem{zhang2013hybrid}
Yanbo Zhang, Hao Yan, Xun Jia, Jian Yang, Steve~B Jiang, and Xuanqin Mou.
\newblock A hybrid metal artifact reduction algorithm for x-ray ct.
\newblock {\em Medical physics}, 40(4):041910, 2013.

\bibitem{zhang2014metal}
Yanbo Zhang and Xuanqian Mou.
\newblock Metal artifact reduction based on the combined prior image.
\newblock {\em arXiv preprint arXiv:1408.5198}, 2014.

\bibitem{Wang2013}
Jun Wang, Shijie Wang, Yang Chen, Jiasong Wu, Jean~Louis Coatrieux, and Limin
  Luo.
\newblock {Metal artifact reduction in CT using fusion based prior image}.
\newblock {\em Medical Physics}, 40(8):1--8, 2013.

\bibitem{zhang2018convolutional}
Yanbo Zhang and Hengyong Yu.
\newblock Convolutional neural network based metal artifact reduction in x-ray
  computed tomography.
\newblock {\em IEEE transactions on medical imaging}, 37(6):1370--1381, 2018.

\bibitem{Lin2019}
Wei-An Lin, Haofu Liao, Cheng Peng, Xiaohang Sun, Jingdan Zhang, Jiebo Luo,
  Rama Chellappa, and Shaohua~Kevin Zhou.
\newblock {DuDoNet: Dual Domain Network for CT Metal Artifact Reduction}.
\newblock CVPR, 2019.

\bibitem{Lyu2020}
Yuanyuan Lyu, Wei-An Lin, Jingjing Lu, and S.~Kevin Zhou.
\newblock {DuDoNet++: Encoding mask projection to reduce CT metal artifacts}.
\newblock jan 2020.

\bibitem{yu2020deep}
Lequan Yu, Zhicheng Zhang, Xiaomeng Li, and Lei Xing.
\newblock Deep sinogram completion with image prior for metal artifact
  reduction in ct images.
\newblock {\em IEEE Transactions on Medical Imaging}, 40(1):228--238, 2020.

\bibitem{wang2021dan}
Tao Wang, Wenjun Xia, Yongqiang Huang, Huaiqiang Sun, Yan Liu, Hu~Chen, Jiliu
  Zhou, and Yi~Zhang.
\newblock Dan-net: Dual-domain adaptive-scaling non-local network for ct metal
  artifact reduction.
\newblock {\em Physics in Medicine \& Biology}, 66(15):155009, 2021.

\bibitem{Yu2021}
Lequan Yu, Zhicheng Zhang, Xiaomeng Li, Hongyi Ren, Wei Zhao, and Lei Xing.
\newblock {Metal artifact reduction in 2D CT images with self-supervised
  cross-domain learning}.
\newblock {\em Physics in Medicine \& Biology}, 66(17):175003, aug 2021.

\bibitem{meilinger2011projective}
Manuel Meilinger, Elmar~W Lang, Christian Schmidgunst, and Oliver Sch{\"u}tz.
\newblock Projective segmentation of metal implants in cone beam computed
  tomographic images.
\newblock In {\em 2011 7th International Symposium on Image and Signal
  Processing and Analysis (ISPA)}, pages 507--512. IEEE, 2011.

\bibitem{hegazy2019u}
Mohamed~AA Hegazy, Myung~Hye Cho, Min~Hyoung Cho, and Soo~Yeol Lee.
\newblock U-net based metal segmentation on projection domain for metal
  artifact reduction in dental ct.
\newblock {\em Biomedical engineering letters}, 9:375--385, 2019.

\bibitem{gottschalk2021view}
Tristan~M Gottschalk, Andreas Maier, Florian Kordon, and Bj{\"o}rn~W Kreher.
\newblock View-consistent metal segmentation in the projection domain for metal
  artifact reduction in cbct--an investigation of potential improvement.
\newblock {\em arXiv preprint arXiv:2112.02101}, 2021.

\bibitem{zhu2021ct}
Yulin Zhu, Xiaokun Liang, Lei Deng, Chenglong Zhang, Xuanru Zhou, Yaoqin Xie,
  and Huailing Zhang.
\newblock Ct metal artifact correction assisted by the deep learning-based
  metal segmentation on the projection domain.
\newblock In {\em 2021 IEEE International Conference on Medical Imaging Physics
  and Engineering (ICMIPE)}, pages 1--10. IEEE, 2021.

\bibitem{Meijering2004}
E.~Meijering, M.~Jacob, J.~C.F. Sarria, P.~Steiner, H.~Hirling, and M.~Unser.
\newblock {Neurite tracing in fluorescence microscopy images using ridge
  filtering and graph searching: Principles and validation}.
\newblock {\em 2004 2nd IEEE International Symposium on Biomedical Imaging:
  Macro to Nano}, 2(1):1219--1222, 2004.

\bibitem{delaunay1934sphere}
Boris Delaunay et~al.
\newblock Sur la sphere vide.
\newblock {\em Izv. Akad. Nauk SSSR, Otdelenie Matematicheskii i Estestvennyka
  Nauk}, 7(793-800):1--2, 1934.

\bibitem{feldkamp1984practical}
Lee~A Feldkamp, Lloyd~C Davis, and James~W Kress.
\newblock Practical cone-beam algorithm.
\newblock {\em Josa a}, 1(6):612--619, 1984.

\bibitem{wu2022xcist}
Mingye Wu, Paul FitzGerald, Jiayong Zhang, W~Paul Segars, Hengyong Yu, Yongshun
  Xu, and Bruno De~Man.
\newblock Xcist—an open access x-ray/ct simulation toolkit.
\newblock {\em Physics in Medicine \& Biology}, 67(19):194002, 2022.

\bibitem{Segars2010}
W.~P. Segars, G.~Sturgeon, S.~Mendonca, Jason Grimes, and B.~M.~W. Tsui.
\newblock {4D XCAT phantom for multimodality imaging research}.
\newblock {\em Medical Physics}, 37(9):4902--4915, sep 2010.

\bibitem{yu2009metal}
Lifeng Yu, Hua Li, Jan Mueller, James~M Kofler, Xin Liu, Andrew~N Primak,
  Joel~G Fletcher, Luis~S Guimaraes, Thanila Macedo, and Cynthia~H McCollough.
\newblock Metal artifact reduction from reformatted projections for hip
  prostheses in multislice helical computed tomography: techniques and initial
  clinical results.
\newblock {\em Investigative radiology}, 44(11):691, 2009.

\bibitem{Abdoli2011}
Mehrsima Abdoli, Johan~R. {De Jong}, Jan Pruim, Rudi~A.J.O. Dierckx, and Habib
  Zaidi.
\newblock {Reduction of artefacts caused by hip implants in CT-based
  attenuation-corrected PET images using 2-D interpolation of a virtual
  sinogram on an irregular grid}.
\newblock {\em European Journal of Nuclear Medicine and Molecular Imaging},
  38(12):2257--2268, 2011.

\end{thebibliography}

\end{document}